\documentclass{mem}
\usepackage{natbib}\usepackage{txfonts}\usepackage{balance}
\usepackage{graphicx}
\usepackage[a4paper]{hyperref}
\idline{75}{282}
\begin{document}
\def\teff{$T\rm_{eff }$}
\def\kms{$\mathrm {km s}^{-1}$}

\title{
Non-thermal acceleration mechanisms in supernova remnant shells}

\subtitle{}

\author{
A. \,Decourchelle\inst{1} }

  \offprints{A. Decourchelle}

\institute{Laboratoire AIM, CEA/DSM - CNRS - Universit\'e Paris Diderot, DAPNIA/Service d'Astrophysique, B\^at. 709, CEA-Saclay, F-91191 Gif-sur- Yvette C\'edex, France
\email{anne.decourchelle@cea.fr}
}

\authorrunning{Decourchelle }

\titlerunning{Acceleration mechanisms in SNRs}

\abstract{ A review of the main issues in the field of particle acceleration in Supernova Remnants is provided in the context of future X-ray observations with {\it Simbol-X}.  After a summary of the nonthermal acceleration mechanisms at work, I briefly review the observations of supernova remnants in hard X-rays and in gamma rays.  Open issues are discussed in this framework.

}
\maketitle{}

\section{Introduction}

Supernova remnants (SNRs) are expected to be the main source of cosmic rays with energies up to the knee at $\simeq 3~10^{15}$~eV \citep[see for a discussion][]{drury01}. The ejection of high velocity material during the explosion gives rise to a shock that both heats the ambient medium to millions degrees and accelerates particles.

While the millions degrees gas emits thermal X-rays below 10 keV, the accelerated charged particles produce non-thermal emission over the entire electromagnetic spectrum through four main emission mechanisms. First, the accelerated electrons produce synchrotron emission when spiraling in a magnetic field. This emission, extending from the radio to X-ray domains, corresponds to a population of GeV and TeV accelerated electrons, respectively. The population of TeV electrons, observed in X-ray synchrotron, gives also rise to Inverse Compton emission: they communicate a large part of their energy to low energy photons (from the Cosmic Microwave Background or from the local optical and infrared environment), which then reach gamma ray energies.
Non-thermal bremsstrahlung is produced by the population of supra-thermal and relativistic electrons when their trajectory is deflected by the electrical field of positive ions: their emission extends from hard X-rays to gamma rays. Finally, the interaction of accelerated protons with those of the ambient medium produces pions. The neutral $\pi^0$ decays almost instantaneously and produces gamma rays in the GeV - TeV range. 

Supernova remnants are well known to be sources of radio synchrotron emission. The first suggestion of such an association dates from 1954 \citep{hanbury54}, and was rapidly confirmed by \citet{baldwin57}. The observed radio shell morphology supports the view that particles (notably electrons) have been accelerated to GeV energies at the shock of SNRs. 

\section{Previous X-ray observations}

While radio synchrotron emission is observed in most SNRs (in 203 over the 217 observed Galactic SNRs, \citealt{green04}), X-ray synchrotron emission is observed only in a few remnants up to now. 
The first evidence of such emission in a shell-like supernova remnant was obtained in SN1006 with {\it ASCA} \citep{koyama95}. {\it ASCA} was the first X-ray satellite able to resolve the spectra of the two bright bilateral rims of SN 1006. This provided the first proof that shocks in supernova remnants were able to accelerate particles (electrons) up to TeV energies. This discovery was followed by the detection with {\it ASCA} of X-ray synchrotron emission in RX J1713.7-3946 \citep{koyama97,slane99}, Vela Junior \citep{slane01}, and  RCW 86 \citep{bamba00}. Since then, these sources have been observed in X-rays in much more details with {\it XMM-Newton} and {\it Chandra} \citep{uchiyama03,cassam04b,iyudin05,bamba05b,vink06}, and in the gamma rays with the {\it H.E.S.S.} and {\it CANGAROO-II} Cherenkov telescopes \citep{katagiri05,aharonian06,aharonian07b}. 
Apart from these synchrotron-dominated shell SNRs, X-ray synchrotron emission has been detected as well in the younger ejecta-dominated SNRs, mainly in Tycho \citep{hwang02}, Cas A \citep{vink03} and Kepler \citep{cassam04a, reynolds07}. For Cas A, there is also evidence for TeV emission \citep{aharonian01}. Above 10 keV, a number of young SNRs show a hard X-ray tail observed with {\it RXTE}, {\it Beppo-Sax} and {\it INTEGRAL}: Cas A \citep{allen97,renaud06}, Tycho, Kepler, SN 1006 and RCW 86 \citep{allen99}. However, as no spectro-imagery is available above 10 keV, the nature of the emission there is still unknown. Different origins may be invoked: Thermal emission from a very high temperature shocked ambient interstellar medium ? Synchrotron emission from relativistic electrons accelerated at the forward shock ? Non-thermal bremsstrahlung from suprathermal electrons accelerated at the contact discontinuity (interface between shocked ejecta and shocked ambient medium) ? 

Spatially resolved spectroscopy above 10 keV is required to observe, to distinguish and to quantify the different contributions of the non-thermal X-ray emission in SNRs. We need {\it Simbol-X} to study the nature of the emission and to answer a number of pending questions on shock physics and particle acceleration.

\section{Shock physics and particle acceleration} 

\subsection{Shock physics}

SNRs produce very high Mach number shocks, that propagate generally in a low density ambient medium ($<1$~cm$^{-3}$). In such conditions, collective plasma processes play a non-negligible role compared to Coulomb collisions \citep{ghavamian07} and this directly affects the degree of temperature equilibration at the shock between the electrons and the ions. 

Quantifying with {\it Simbol-X} the contribution of the thermal emission above 10 keV provides strong constraints on the presence of a high electronic temperature in the shocked ambient medium. If the thermal emission appears to be important, this will impose an efficient heating of the electrons at the shock through collision-less processes, as well as constraints on the supernova kinetic energy.
However, the faint level of thermal emission observed below 10 keV in Tycho \citep{cassam07} and SN 1006 \citep{acero07} does not favor a thermal origin of the hard X-ray emission.

\subsection{Electron acceleration}

Electrons are only about $2\,\%$ of cosmic rays \citep{ferriere01}, but they can reveal a lot on the mechanism of diffusive shock acceleration, as they are accelerated like protons. 

\subsubsection{What is the maximum energy of accelerated particles ?}

A key question for understanding particle acceleration in SNRs lies in knowing the maximum energy of accelerated particles. This can be obtained through the measurement of the cut-off frequency of the synchrotron emission, observable in X-rays, and knowing the magnetic field. Typical values are in the order of 80 TeV assuming a $10\,\mu$G magnetic field \citep{reynolds99}.

Another important issue on particle acceleration is its dependence to the magnetic field orientation. In SN 1006, very strong azimuthal variations of the maximum energy of electrons have been observed with  {\it XMM-Newton} \citep{rothenflug04}, whose amplitude cannot be explained by variations of the magnetic compression alone as proposed by \citet{reynolds96}.  As detailed in \citet{rothenflug04}, the very low synchrotron brightness in the interior of SN 1006 favors a geometry, where the bright non-thermal limbs are polar caps rather than an equatorial belt. In addition, the maximum energy of accelerated particles is higher at the bright limbs than elsewhere. If the magnetic field is amplified at the limbs, the maximum energy is certainly much larger there ($> 1000$~TeV) than outside the limbs (E~$\simeq 25$~TeV if B~$\simeq 10~\mu$G).
The X-ray synchrotron geometry of SN 1006 favors cosmic-ray acceleration where the magnetic field was originally parallel to the shock speed.

There are thus two main objectives on this issue well designed for Simbol-X:\\
- to get a precise determination of the cut-off energy using the full extent of the synchrotron spectrum above 10 keV. \\
- to measure the azimuthal variations of the maximum energy using spatially resolved spectroscopy including energies above 10 keV. 

\subsubsection{How large is the magnetic field ? }

The comparison of the radio and X-ray synchrotron morphologies provide constraints on the downstream magnetic field. In a number of SNRs, the X-ray synchrotron emission exhibits a filamentary emission just behind the blast wave. This is observed in historical \citep{hwang02,vink03,bamba05a} and in synchrotron-dominated \citep{uchiyama03,bamba05b} remnants.

Two explanations have been proposed to interpret that: either the magnetic field is large enough ($\simeq 100~\mu$G) to induce strong radiative losses in the high energy electrons \citep{vink03,ballet06}, or the magnetic field is damped at the shock \citep{pohl05}.

These models predict distinct morphology and spectral shape in X-rays and radio \citep{cassam07}, but the spectrum must be observed over a broad band extending to hard X-rays to tell one from the other. While the expected spatial resolution of Simbol-X is not relevant for studying arcseconds wide filaments, its sensitivity above 10 keV is perfectly suited for constraining the shape of the hard X-ray spectrum in the post-shock region.

\subsection{Proton acceleration}

Is there observational evidence for ion acceleration in SNRs ? How efficient is cosmic-ray acceleration ? What fraction of the shock energy can be tapped by the cosmic rays ? To answer these questions, we can search both for a possible modification of the hydrodynamics and for a direct spectral signature of the accelerated protons.

\subsubsection{Back-reaction of the accelerated particle on the shock and hydrodynamics }

For an efficient ion injection, a large fraction of the shock energy go into accelerated particles. The back-reaction of the accelerated particles modifies the shock structure \citep{ellison00}. The overall shock compression ratio gets larger (above a value of 4) and the post-shock temperature gets lower. Such a case was observed in the remnant 1E 0102.2-7219 in the Small Magellanic Cloud. With an observed shock velocity of about 6200 km/s, the expected mean post-shock temperature is 45 keV. However, the electronic temperature measured with {\it Chandra} is less than 1 keV. Even in the case of complete non-equipartition between the electrons and the ions, namely $T_e/T_i = 1/1836$ at the shock, the Coulomb collisions predict a temperature larger than 2.5 keV. This is interpreted as a signature of efficient proton acceleration at the shock \citep{hughes00}. In case of efficient proton acceleration, not only the shock structure is modified, but also the overall hydrodynamics giving rise to a narrower shocked region \citep{decourchelle00, ellison04}. This morphological feature of proton acceleration is observable in X-rays through the measurement of the ratio between the shock and contact discontinuity radii. In Tycho and Kepler's SNRs, this radii ratio is observed to be less than 1.1, while models without particle acceleration predict radius ratio of $\simeq 1.18$ \citep{decourchelle05}. Using a deep {\it Chandra} observation of Tycho's SNR, \citet{warren05} provided a measurement of the radii ratio as a function of azimuth and confirmed a smaller radii ratio than expected in the test-particle case, featuring efficient acceleration of protons at the shock. One of the uncertainty in the measurement arises from the determination of the contact discontinuity radius which is affected by the development of Rayleigh-Taylor instabilities \citep{blondin01}. This complicates the interpretation of the observed radii ratio.

\subsubsection{Pion decay and TeV observations }

A direct signature of accelerated protons is expected through pion decay emission in the GeV-TeV gamma ray range. However at these energies, a contribution of the Inverse Compton emission is expected as well and requires to be correctly estimated for a proper determination of contribution of the pion decay emission. This can be done by characterizing the population of TeV accelerated electrons responsible for the Inverse Compton emission through their synchrotron emission observed in the X-ray range and by estimating the magnetic field intensity (see sect. 3.2.2). Combining X-ray and gamma-ray observations is thus a fruitful way to estimate the pion decay emission in the TeV energy range.
Ultimately, an unequivocal signature of accelerated protons is expected through gamma ray observations of the spectral feature associated to the pion decay. {\it GLAST}, to be launched early 2008, is expected to provide such observations.

\subsection{Acceleration by secondary shocks}

As discussed in section 2., a number of young SNRs do exhibit a hard X-ray tail \citep{allen99}. However, its origin is still unknown, due to the absence of spatially resolved spectroscopy above 10 keV. Only Cas A supernova remnant has been mapped above 10 keV, thanks to {\it XMM-Newton} \citep{bleeker01}. Its hard X-ray tail observed up to 40 keV was first interpreted as synchrotron emission \citep{allen97}. However, the {\it XMM-Newton} image in the 8-15 keV band revealed that the emission was mostly coming from the interface between the shocked ejecta and the shocked ambient medium, and not from the main blast wave. This finding is inconsistent with a synchrotron interpretation as it requires to have high energy TeV electrons in the regions which have been shocked at earliest in the history of the remnant. This is impossible as the losses affecting the TeV electrons are very efficient and confine them close to the shock. \citet{vink03} proposed that the emission at the interface was non-thermal bremsstrahlung  due to particle acceleration at secondary shocks. The origin of these hard X-ray tails is still an open issue, for which {\it Simbol-X} is awaited.

\section{Contributions of SIMBOL-X}

\begin{figure}[t!]
\resizebox{\hsize}{!}{\includegraphics[clip=true]{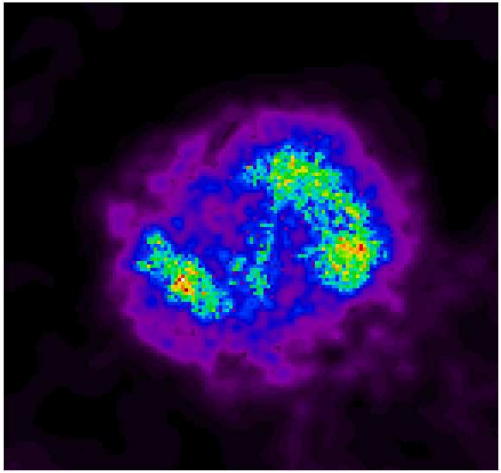}}
\caption{\footnotesize Simulated {\it Simbol-X} image of Cas A in 30 ks for energies above 20 keV. The field of view is $10 \times 10$~arcmin$^2$.}
\label{im_casa}
\end{figure}
%
\begin{figure}[t!]
\resizebox{\hsize}{!}{\includegraphics[clip=true]{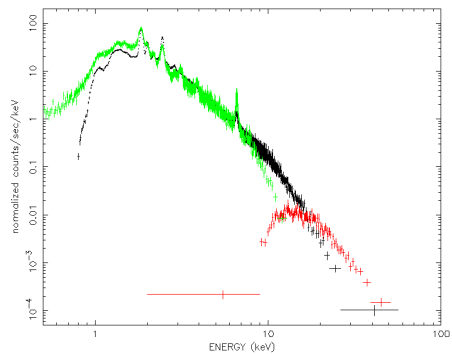}}
\caption{\footnotesize Simulated {\it Simbol-X} spectrum of a region of 1 arcmin$^2$ in the west of Cas A for 30 ks observation. In green the {\it XMM-Newton} EPIC PN spectrum, in black and red the {\it Simbol-X} MPD and CZT spectra, respectively.}
\label{spec_casa}
\end{figure}
%
\begin{figure}[t!]
\resizebox{\hsize}{!}{\includegraphics[clip=true]{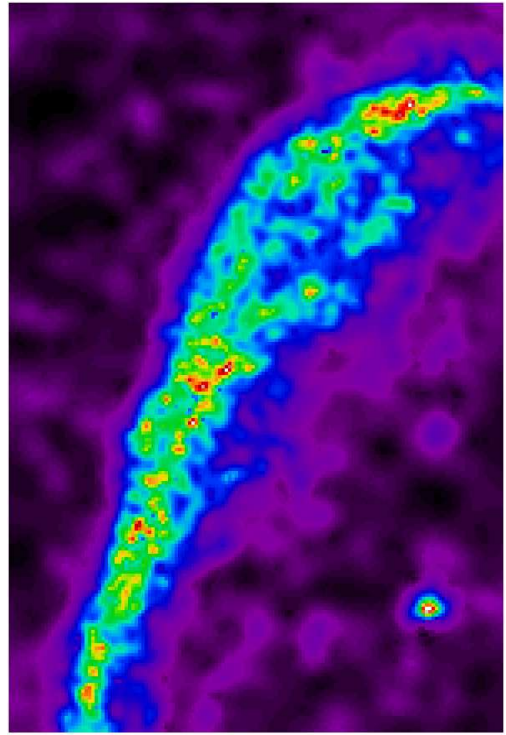}}
\caption{\footnotesize Simulated {\it Simbol-X} image above 10 keV of the north-eastern synchrotron rim in SN 1006 for a 30 ks observation. The field of view is $10 \times 10$~arcmin$^2$.}
\label{im_sn1006} 
\vspace{-0.5cm}
\end{figure}

In the field of supernova remnants, the major expected contributions of {\it Simbol-X}  concern particle acceleration. It will be the first instrument with a full spectro-imaging capability above 10 keV, and a broad energy range covering at a good spatial resolution both the thermal and the non-thermal regimes. It is only with such a spectro-imaging capability that the contribution of each potential emission processus can be distinguished and quantified: thermal emission from a high temperature plasma, nonthermal bremsstrahlung from a population of suprathermal electrons, or  synchrotron emission from TeV accelerated electrons. With its broad energy range, {\it Simbol-X}  is well suited to investigate the coupling between thermal and non-thermal populations through the back-reaction of accelerated protons on the shock structure and on the overall hydrodynamics.
With its spatially resolved spectroscopy capability above 10 keV, {\it Simbol-X}  can study the physics of the acceleration of electrons, the physics at the interface between the ejecta and the ambient medium, as well as the physics at the shock through the heating of the electrons. The observation of the $^{44}$Ti gamma ray lines is also an objective for {\it Simbol-X}, as discussed in these proceedings by Renaud et al.

The best potential SNR sources are : \\ 
- the young ejecta-dominated Galactic SNRs. They are relatively bright sources, well adapted to the field of view. They do exhibit thermal and non-thermal emission, as well as a pronounced hard X-ray tail (see Cas A, Figs. 1 and 2). \\
- the specific sites of particle acceleration in synchrotron-dominated SNRs. They are relatively bright sources of hard X-ray emission (see SN 1006 in Fig. 3), showing strong spectral variations (below 10 keV).


\bibliographystyle{aa} 
\bibliography{decourchelle} 

\begin{thebibliography}{43}
\expandafter\ifx\csname natexlab\endcsname\relax\def\natexlab#1{#1}\fi

\bibitem[{{Ac\'ero} {et~al.}(2007){Ac\'ero}, {Ballet}, \&
  {Decourchelle}}]{acero07}
{Ac\'ero}, F., {Ballet}, J., \& {Decourchelle}, A. 2007, \aap, sub.

\bibitem[{{Aharonian} {et~al.}(2001){Aharonian}, {Akhperjanian}, {Barrio},
  {Bernl{\"o}hr}, {B{\"o}rst}, {Bojahr}, {Bolz}, {Contreras}, {Cortina},
  {Denninghoff}, {Fonseca}, {Gonzalez}, {G{\"o}tting}, {Heinzelmann},
  {Hermann}, {Heusler}, {Hofmann}, {Horns}, {Ibarra}, {Iserlohe}, {Jung},
  {Kankanyan}, {Kestel}, {Kettler}, {Kohnle}, {Konopelko}, {Kornmeyer},
  {Kranich}, {Krawczynski}, {Lampeitl}, {Lopez}, {Lorenz}, {Lucarelli},
  {Magnussen}, {Mang}, {Meyer}, {Mirzoyan}, {Moralejo}, {Ona}, {Padilla},
  {Panter}, {Plaga}, {Plyasheshnikov}, {Prahl}, {P{\"u}hlhofer}, {Rauterberg},
  {R{\"o}hring}, {Rhode}, {Rowell}, {Sahakian}, {Samorski}, {Schilling},
  {Schr{\"o}der}, {Siems}, {Stamm}, {Tluczykont}, {V{\"o}lk}, {Wiedner}, \&
  {Wittek}}]{aharonian01}
{Aharonian}, F., {Akhperjanian}, A., {Barrio}, J., {et~al.} 2001, \aap, 370,
  112

\bibitem[{{Aharonian} {et~al.}(2007){Aharonian}, {Akhperjanian}, {Bazer-Bachi},
  {Beilicke}, {Benbow}, {Berge}, {Bernl{\"o}hr}, {Boisson}, {Bolz}, {Borrel},
  {Braun}, {Brown}, {B{\"u}hler}, {B{\"u}sching}, {Carrigan}, {Chadwick},
  {Chounet}, {Coignet}, {Cornils}, {Costamante}, {Degrange}, {Dickinson},
  {Djannati-Ata{\"\i}}, {Drury}, {Dubus}, {Egberts}, {Emmanoulopoulos},
  {Espigat}, {Feinstein}, {Ferrero}, {Fiasson}, {Filipovic}, {Fontaine},
  {Fukui}, {Funk}, {Funk}, {F{\"u}{\ss}ling}, {Gallant}, {Giebels},
  {Glicenstein}, {Goret}, {Hadjichristidis}, {Hauser}, {Hauser}, {Heinzelmann},
  {Henri}, {Hermann}, {Hinton}, {Hiraga}, {Hoffmann}, {Hofmann}, {Holleran},
  {Hoppe}, {Horns}, {Ishisaki}, {Jacholkowska}, {de Jager}, {Kendziorra},
  {Kerschhaggl}, {Kh{\'e}lifi}, {Komin}, {Konopelko}, {Kosack}, {Lamanna},
  {Latham}, {Le Gallou}, {Lemi{\`e}re}, {Lemoine-Goumard}, {Lohse}, {Martin},
  {Martineau-Huynh}, {Marcowith}, {Masterson}, {Maurin}, {McComb}, {Moulin},
  {Moriguchi}, {de Naurois}, {Nedbal}, {Nolan}, {Noutsos}, {Orford}, {Osborne},
  {Ouchrif}, {Panter}, {Pelletier}, {Pita}, {P{\"u}hlhofer}, {Punch},
  {Ranchon}, {Raubenheimer}, {Raue}, {Rayner}, {Reimer}, {Ripken}, {Rob},
  {Rolland}, {Rosier-Lees}, {Rowell}, {Sahakian}, {Santangelo}, {Saug{\'e}},
  {Schlenker}, {Schlickeiser}, {Schr{\"o}der}, {Schwanke}, {Schwarzburg},
  {Schwemmer}, {Shalchi}, {Sol}, {Spangler}, {Spanier}, {Steenkamp},
  {Stegmann}, {Superina}, {Tam}, {Tavernet}, {Terrier}, {Tluczykont}, {van
  Eldik}, {Vasileiadis}, {Venter}, {Vialle}, {Vincent}, {V{\"o}lk}, {Wagner},
  \& {Ward}}]{aharonian07b}
{Aharonian}, F., {Akhperjanian}, A.~G., {Bazer-Bachi}, A.~R., {et~al.} 2007,
  \apj, 661, 236

\bibitem[{Aharonian \& al.(2006)}]{aharonian06}
Aharonian, F.~A. \& al. 2006, A\&A, 449, 223

\bibitem[{Allen {et~al.}(1999)Allen, Gotthelf, \& Petre}]{allen99}
Allen, G.~E., Gotthelf, E., \& Petre, R. 1999, in International Cosmic Ray
  Conference, 480--+

\bibitem[{Allen {et~al.}(1997)Allen, Keohane, Gotthelf, Petre, Jahoda,
  Rothschild, Lingenfelter, Heindl, Marsden, Gruber, Pelling, \&
  Blanco}]{allen97}
Allen, G.~E., Keohane, J.~W., Gotthelf, E.~V., {et~al.} 1997, ApJL, 487, L97+

\bibitem[{{Baldwin} \& {Edge}(1957)}]{baldwin57}
{Baldwin}, J.~E. \& {Edge}, D.~O. 1957, The Observatory, 77, 139

\bibitem[{{Ballet}(2006)}]{ballet06}
{Ballet}, J. 2006, Advances in Space Research, 37, 1902

\bibitem[{{Bamba} {et~al.}(2000){Bamba}, {Koyama}, \& {Tomida}}]{bamba00}
{Bamba}, A., {Koyama}, K., \& {Tomida}, H. 2000, \pasj, 52, 1157

\bibitem[{{Bamba} {et~al.}(2005){Bamba}, {Yamazaki}, \& {Hiraga}}]{bamba05b}
{Bamba}, A., {Yamazaki}, R., \& {Hiraga}, J.~S. 2005, \apj, 632, 294

\bibitem[{Bamba {et~al.}(2005)Bamba, Yamazaki, Yoshida, Terasawa, \&
  Koyama}]{bamba05a}
Bamba, A., Yamazaki, R., Yoshida, T., Terasawa, T., \& Koyama, K. 2005, ApJ,
  621, 793

\bibitem[{{Bleeker} {et~al.}(2001){Bleeker}, {Willingale}, {van der Heyden},
  {Dennerl}, {Kaastra}, {Aschenbach}, \& {Vink}}]{bleeker01}
{Bleeker}, J.~A.~M., {Willingale}, R., {van der Heyden}, K., {et~al.} 2001,
  \aap, 365, L225

\bibitem[{Blondin \& Ellison(2001)}]{blondin01}
Blondin, J.~M. \& Ellison, D.~C. 2001, \apj, 560, 244

\bibitem[{Cassam-Chena{\"\i} {et~al.}(2004{\natexlab{a}})Cassam-Chena{\"\i},
  Decourchelle, Ballet, Hwang, Hughes, \& Petre}]{cassam04a}
Cassam-Chena{\"\i}, G., Decourchelle, A., Ballet, J., {et~al.}
  2004{\natexlab{a}}, A\&A, 414, 545

\bibitem[{Cassam-Chena{\"\i} {et~al.}(2004{\natexlab{b}})Cassam-Chena{\"\i},
  Decourchelle, Ballet, Sauvageot, Dubner, \& Giacani}]{cassam04b}
Cassam-Chena{\"\i}, G., Decourchelle, A., Ballet, J., {et~al.}
  2004{\natexlab{b}}, A\&A, 427, 199

\bibitem[{{Cassam-Chena{\"\i}} {et~al.}(2007){Cassam-Chena{\"\i}}, {Hughes},
  {Ballet}, \& {Decourchelle}}]{cassam07}
{Cassam-Chena{\"\i}}, G., {Hughes}, J.~P., {Ballet}, J., \& {Decourchelle}, A.
  2007, \apj, 665, 315

\bibitem[{{Decourchelle}(2005)}]{decourchelle05}
{Decourchelle}, A. 2005, in X-Ray and Radio Connections (eds. L.O. Sjouwerman
  and K.K Dyer) Published electronically by NRAO,
  http://www.aoc.nrao.edu/events/xraydio Held 3-6 February 2004 in Santa Fe,
  New Mexico, USA, (E4.02) 10 pages

\bibitem[{Decourchelle {et~al.}(2000)Decourchelle, Ellison, \&
  Ballet}]{decourchelle00}
Decourchelle, A., Ellison, D.~C., \& Ballet, J. 2000, ApJ, 543, L57

\bibitem[{Drury {et~al.}(2001)Drury, Ellison, Aharonian, Berezhko, Bykov,
  Decourchelle, Diehl, Meynet, Parizot, Raymond, Reynolds, \&
  Spangler}]{drury01}
Drury, L.~O., Ellison, D.~E., Aharonian, F.~A., {et~al.} 2001, Space Science
  Reviews, 99, 329

\bibitem[{{Ellison}(2000)}]{ellison00}
{Ellison}, D.~C. 2000, in American Institute of Physics Conference Series, Vol.
  528, Acceleration and Transport of Energetic Particles Observed in the
  Heliosphere, ed. R.~A. {Mewaldt}, J.~R. {Jokipii}, M.~A. {Lee},
  E.~{M{\"o}bius}, \& T.~H. {Zurbuchen}, 383--+

\bibitem[{Ellison {et~al.}(2004)Ellison, Decourchelle, \& Ballet}]{ellison04}
Ellison, D.~C., Decourchelle, A., \& Ballet, J. 2004, A\&A, 413, 189

\bibitem[{Ferri{\`e}re(2001)}]{ferriere01}
Ferri{\`e}re, K.~M. 2001, Reviews of Modern Physics, 73, 1031

\bibitem[{Ghavamian {et~al.}(2007)Ghavamian, Laming, \& Rakowski}]{ghavamian07}
Ghavamian, P., Laming, J.~M., \& Rakowski, C.~E. 2007, ApJL, 654, L69

\bibitem[{{Green}(2004)}]{green04}
{Green}, D.~A. 2004, Bulletin of the Astronomical Society of India, 32, 335

\bibitem[{{Hanbury Brown}(1954)}]{hanbury54}
{Hanbury Brown}, R. 1954, The Observatory, 74, 185

\bibitem[{Hughes {et~al.}(2000)Hughes, Rakowski, \& Decourchelle}]{hughes00}
Hughes, J.~P., Rakowski, C.~E., \& Decourchelle, A. 2000, ApJ, 543, L61

\bibitem[{Hwang {et~al.}(2002)Hwang, Decourchelle, Holt, \& Petre}]{hwang02}
Hwang, U., Decourchelle, A., Holt, S.~S., \& Petre, R. 2002, ApJ, 581, 1101

\bibitem[{{Iyudin} {et~al.}(2005){Iyudin}, {Aschenbach}, {Becker}, {Dennerl},
  \& {Haberl}}]{iyudin05}
{Iyudin}, A.~F., {Aschenbach}, B., {Becker}, W., {Dennerl}, K., \& {Haberl}, F.
  2005, \aap, 429, 225

\bibitem[{{Katagiri} {et~al.}(2005){Katagiri}, {Enomoto}, {Ksenofontov},
  {Mori}, {Adachi}, {Asahara}, {Bicknell}, {Clay}, {Doi}, {Edwards}, {Gunji},
  {Hara}, {Hara}, {Hattori}, {Hayashi}, {Itoh}, {Kabuki}, {Kajino}, {Kawachi},
  {Kifune}, {Kiuchi}, {Kubo}, {Kurihara}, {Kurosaka}, {Kushida}, {Matsubara},
  {Miyashita}, {Mizumoto}, {Muraishi}, {Muraki}, {Naito}, {Nakamori}, {Nakase},
  {Nishida}, {Nishijima}, {Ohishi}, {Okumura}, {Patterson}, {Protheroe},
  {Sakamoto}, {Sakamoto}, {Swaby}, {Tanimori}, {Tanimura}, {Thornton},
  {Tsuchiya}, {Watanabe}, {Yamaoka}, {Yanagita}, {Yoshida}, \&
  {Yoshikoshi}}]{katagiri05}
{Katagiri}, H., {Enomoto}, R., {Ksenofontov}, L.~T., {et~al.} 2005, \apjl, 619,
  L163

\bibitem[{{Koyama} {et~al.}(1997){Koyama}, {Kinugasa}, {Matsuzaki},
  {Nishiuchi}, {Sugizaki}, {Torii}, {Yamauchi}, \& {Aschenbach}}]{koyama97}
{Koyama}, K., {Kinugasa}, K., {Matsuzaki}, K., {et~al.} 1997, \pasj, 49, L7

\bibitem[{{Koyama} {et~al.}(1995){Koyama}, {Petre}, {Gotthelf}, {Hwang},
  {Matsuura}, {Ozaki}, \& {Holt}}]{koyama95}
{Koyama}, K., {Petre}, R., {Gotthelf}, E.~V., {et~al.} 1995, \nat, 378, 255

\bibitem[{Pohl {et~al.}(2005)Pohl, Yan, \& Lazarian}]{pohl05}
Pohl, M., Yan, H., \& Lazarian, A. 2005, ApJ, 626, L101

\bibitem[{Renaud {et~al.}(2006)Renaud, Vink, Decourchelle, Lebrun, den Hartog,
  Terrier, Couvreur, Kn{\"o}dlseder, Martin, Prantzos, Bykov, \&
  Bloemen}]{renaud06}
Renaud, M., Vink, J., Decourchelle, A., {et~al.} 2006, ApJ, 647, L41

\bibitem[{{Reynolds}(1996)}]{reynolds96}
{Reynolds}, S.~P. 1996, \apjl, 459, L13+

\bibitem[{{Reynolds} {et~al.}(2007){Reynolds}, {Borkowski}, {Hwang}, {Hughes},
  {Badenes}, {Laming}, \& {Blondin}}]{reynolds07}
{Reynolds}, S.~P., {Borkowski}, K.~J., {Hwang}, U., {et~al.} 2007, ApJL, 708

\bibitem[{Reynolds \& Keohane(1999)}]{reynolds99}
Reynolds, S.~P. \& Keohane, J.~W. 1999, \apj, 525, 368

\bibitem[{Rothenflug {et~al.}(2004)Rothenflug, Ballet, Dubner, Giacani,
  Decourchelle, \& Ferrando}]{rothenflug04}
Rothenflug, R., Ballet, J., Dubner, G., {et~al.} 2004, A\&A, 425, 121

\bibitem[{{Slane} {et~al.}(1999){Slane}, {Gaensler}, {Dame}, {Hughes},
  {Plucinsky}, \& {Green}}]{slane99}
{Slane}, P., {Gaensler}, B.~M., {Dame}, T.~M., {et~al.} 1999, \apj, 525, 357

\bibitem[{{Slane} {et~al.}(2001){Slane}, {Hughes}, {Edgar}, {Plucinsky},
  {Miyata}, {Tsunemi}, \& {Aschenbach}}]{slane01}
{Slane}, P., {Hughes}, J.~P., {Edgar}, R.~J., {et~al.} 2001, \apj, 548, 814

\bibitem[{{Uchiyama} {et~al.}(2003){Uchiyama}, {Aharonian}, \&
  {Takahashi}}]{uchiyama03}
{Uchiyama}, Y., {Aharonian}, F.~A., \& {Takahashi}, T. 2003, \aap, 400, 567

\bibitem[{Vink {et~al.}(2006)Vink, Bleeker, van~der Heyden, Bykov, Bamba, \&
  Yamazaki}]{vink06}
Vink, J., Bleeker, J., van~der Heyden, K., {et~al.} 2006, ApJL, 648, L33

\bibitem[{Vink \& Laming(2003)}]{vink03}
Vink, J. \& Laming, J.~M. 2003, ApJ, 584, 758

\bibitem[{Warren {et~al.}(2005)Warren, Hughes, Badenes, Ghavamian, McKee,
  Moffett, Plucinsky, Rakowski, Reynoso, \& Slane}]{warren05}
Warren, J.~S., Hughes, J.~P., Badenes, C., {et~al.} 2005, \apj, 634, 376

\end{thebibliography}

\end{document}